# Revealing the role of orbital magnetism in ultrafast laser-induced demagnetization in multisublattice metallic magnets by terahertz spectroscopy

T. J. Huisman[1], R. V. Mikhaylovskiy[1], A. Tsukamoto[2], Th. Rasing[1] and A. V. Kimel[1]

[1]*Radboud University Nijmegen, Institute for Molecules and Materials, 6525 AJ Nijmegen, The Netherlands*
[2]*College of Science and Technology, Nihon University, 7-24-1 Funabashi, Chiba, Japan*

**Simultaneous detection of THz emission and transient magneto-optical response is employed to study ultrafast laser-induced magnetization dynamics in multisublattice magnets NdFeCo and GdFeCo amorphous alloys with in-plane magnetic anisotropy. A satisfactory quantitative agreement between the dynamics revealed with the help of these two techniques is obtained for GdFeCo. For NdFeCo the THz emission reveals faster dynamics than the magneto-optical response. This indicates that in addition to spin dynamics of Fe ultrafast laser excitation triggers faster magnetization dynamics of Nd.**

Spin-orbit coupling is a key mechanism exploited by the fast expanding field of spintronics, which ties charge and spin degrees of freedom together [1,2,3]. Besides spin or charge manipulation, one could try to manipulate the orbital degree of freedom. This leads to the intriguing question about the role of orbital magnetization in magnetization dynamics. The seminal observation of subpicosecond demagnetization in ferromagnetic nickel by a laser pulse published almost two decades ago [4] triggered the field of ultrafast magnetization dynamics - a topic that has been continuously fueled by intriguing observations as well as controversies in the scientific community [5-11]. One of the main reasons for the controversies is the lack of an artifact-free technique capable of observing the magnetization dynamics at the subpicosecond timescale.

Most experimental studies of ultrafast magnetism performed so far employ an all-optical pump-probe technique in which the magnetization is probed indirectly via the magneto-optical Faraday or Kerr effect. However it was noticed that at the subpicosecond time-scale the magneto-optical probes are subjective to artifacts [12]. In particular, it was argued that if the temporal behavior of the Kerr ellipticity is different from the one of the Kerr rotation, the dynamics of the magneto-optical signal cannot be directly associated with the true magnetization dynamics [13]. However, the opposite statement is not obviously true and similar behavior of the ellipticity and the rotation cannot be used as a proof that the dynamics of the magneto-optical Kerr effect (MOKE) adequately reflects the magnetization dynamics. Additional complications in the interpretation of time-resolved magneto-optical experiments arise in multi-sublattice magnets, especially when magnetic sublattices possess both spin and orbital magnetization [14-19]. Alternatively, in order to deduce information about subpicosecond magnetization dynamics one can employ the fact that such dynamics is accompanied by the emission of terahertz (THz) electro-magnetic radiation [20-23].

To reveal how orbital magnetization affects ultrafast laser induced demagnetization measurements, we used simultaneous measurements of the MOKE and the electric field of the emitted THz radiation and applied both probes of demagnetization to NdFeCo and GdFeCo alloys. The orbital momenta of Fe and Gd in the GdFeCo alloys are expected to be quenched, while in Nd the magnetization is dominantly determined by its orbital momentum. This makes these samples suitable for studying the role of orbital momentum in ultrafast magnetization dynamics [15,19].

We show that nonlinear optical effects of non-magnetic origin or the inverse spin-Hall effect cannot be responsible for the observed emission of the THz radiation, and thus the emission originates from the ultrafast demagnetization. Assuming that the MOKE represents the dynamics of the net magnetization, we deduce the spectrum of the THz emission which is generated by such magnetization dynamics by solving the Maxwell equations for a thin magnetic film, taking into account the propagation of the THz radiation from the emitter to the detector. Comparing these calculated spectra with the actually measured ones we reveal that there is a very good match between the spectra for a pure Co film, which was used as a reference. While qualitative similarities between the spectra are obtained for GdFeCo alloys, in NdFeCo alloys there are clear discrepancies between the calculated and the actually measured THz spectra. This finding shows that the actual laser-induced magnetization dynamics in NdFeCo alloys is faster and more complicated than time-resolved magneto-optical measurements might reveal, providing indication of the Nd orbital magnetization role in ultrafast demagnetization.

The multisublattice magnetic materials studied in this work are rare-earth transition-metal amorphous alloys: NdFeCo and GdFeCo. The results presented here are for $Nd_{0.2}(Fe_{0.87}Co_{0.13})_{0.8}$ and $Gd_{0.3}(Fe_{0.87}Co_{0.13})_{0.7}$, measured at room temperature. We note that temperature dependent measurements in the range 150K to 300K as well as similar measurements on $Nd_{0.5}(Fe_{0.87}Co_{0.13})_{0.5}$ and $Gd_{0.18}(Fe_{0.87}Co_{0.13})_{0.82}$ show similar results. The GdFeCo alloys are ferrimagnetic and the properties of these alloys are essentially determined by the fact that the Gd spin moments (4*f* and 5*d*) are aligned oppositely to the spin moments of Fe (3*d*) and Co. Importantly, in $Gd_{0.18}(Fe_{0.87}Co_{0.13})_{0.82}$ the FeCo magnetization is larger than the one of Gd at all temperatures, while in $Gd_{0.3}(Fe_{0.87}Co_{0.13})_{0.7}$ the net magnetization is dominated by the Gd sublattice. In the NdFeCo alloy, the orbital momentum of Nd is larger than the spin of Nd and aligned antiparallel with respect to it. Hence, despite the antiferromagnetic coupling of the Fe and Nd spins, the alloy is effectively ferromagnetic. The NdFeCo and GdFeCo alloys were incorporated into a layered structure with different layers to prevent oxidation (capping layer SiN), reduce laser-induced heating (heat sink AlTi) and facilitate the growth (buffer layer SiN). The layered structure is shown in Fig. 1 (a). As a reference we used a 12 nm thick pure Co film deposited on a 0.5 mm thick glass substrate. All the magnetic films have in-plane magnetic anisotropy with coercive fields below 150 Gauss at room temperature.

Our experimental approach uniquely combines THz time-domain spectroscopy (THz-TDS) with an optical pump-probe scheme, as sketched in Fig. 1 (a). An amplified titanium-sapphire laser system is used, producing light pulses with a duration of 50 fs at a repetition rate of 1 kHz and with a center wavelength of 800 nm. The laser beam is divided in three parts: pump, probe and gate. The pump fluence was approximately 1 mJ/cm$^2$. The pump was focused onto an area on the sample with a diameter of approximately 1 mm. THz radiation emitted from the sample is collected and focused onto a ZnTe crystal using two parabolic mirrors. The THz radiation induces birefringence inside the ZnTe crystal due to the electro-optic Pockels effect. By probing this birefringence using the gate pulse and a balanced bridge detection scheme, we are able to reconstruct the electric field of the emitted THz radiation as a function of time. Simultaneously, the probe pulse with a fluence of approximately 10 times less than the pump is focused on the sample to a spot which is approximately twice smaller in diameter than the pump. The angle of incidence of the probe beam is 25 degrees. By measuring the polarization rotation of the reflected probe pulse, information about the ultrafast laser-induced magnetization dynamics is obtained by means of the MOKE. The measurements were performed in magnetic fields up to 1 kG applied in-plane of the samples. Performing the measurements for two polarities of the magnetic field and taking the difference between the measurements, one can deduce signals odd with respect to the field and thus minimize any influence of artifacts of non-magnetic origin. Simultaneous MOKE and THz detection of the laser-induced dynamics minimize potential ambiguities, which may arise when comparing two different experiments.

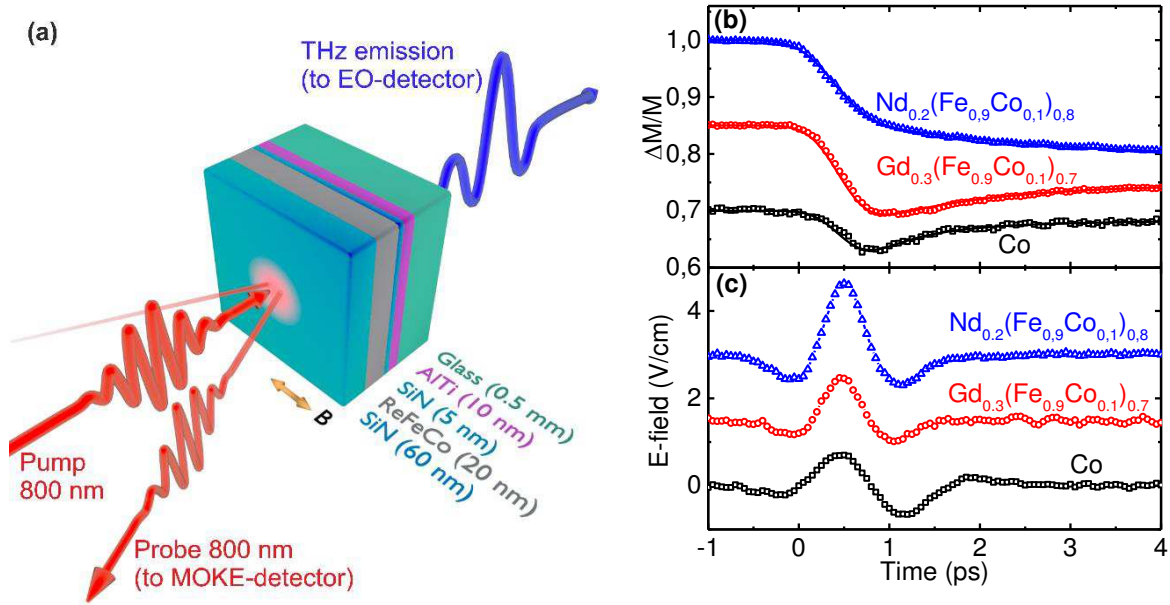

FIG. 1. (color online). (a) Layered structure under study and the scheme of the experiment. Re is rare-earth being either Nd or Gd. The reflected probe pulse is used to measure the MOKE signal. The electric field of the THz emission is detected with the help of electro-optic (EO) detection. A field of +/- 1 kG is applied in-plane. (b) Ultrafast magnetization dynamics as deduced from time-resolved MOKE measurements. The solid lines are fitting functions as described in the supplementary information. (c) Electric field of the emitted THz radiation as a function of time. The position of zero-time is chosen arbitrary so that the position of the peak of the pump pulse is close to zero. The traces are vertically shifted with respect to each other.

Figure 1 (b) shows the results of all-optical pump-probe experiments in which the laser-induced dynamics is probed with the help of the MOKE. Using static MOKE measurements the transient signals were calibrated. The figure shows the data in units of the relative change of the magnetization. NdFeCo shows clearly distinct dynamics from that of GdFeCo and the reference Co-sample.

Figure 1 (c) shows the temporal evolution of the electric field of the emitted radiation for the three samples. With the help of wire-grid polarizers, we found that the electric field of the THz emission from all the samples is linearly polarized, perpendicular to the magnetization. Figure 2 shows the THz waveforms measured for the $Nd_{0.2}(Fe_{0.87}Co_{0.13})_{0.8}$ alloy, which is representative for all the studied samples. The traces clearly change sign when the applied field is reversed. Moreover, measuring the peak amplitude as a function of the applied magnetic field reveals a hysteresis behavior as shown on the inset of Fig. 2. This observation is a clear demonstration of the fact that the electric field of the THz emission is proportional to the magnetization of the magnetic layer.

One can argue that the THz emission is not due to ultrafast laser-induced demagnetization, but has an electric-dipole nature [21] and occurs because of effective symmetry breaking in the studied heterostructure. Such a symmetry breaking is especially efficient at the interfaces of the magnetic film. One of the microscopic realizations of this mechanism is suggested in Ref. [23] showing that the THz emission is generated due to the inverse spin Hall-effect experienced by a spin-polarized current from a magnetic to a non-magnetic layer. To check for emission generated by such a spin current from the magnetic layer, we inverted the sample by turning it around while keeping the applied magnetic field fixed. In this way the direction of a potential spin current between adjacent layers has to reverse sign.

Therefore the sign of the THz radiation which would originate from this current, should be reversed as well. Figure 2 shows the THz waveforms after turning the sample around. Besides different pulse widths and amplitude changes, which are mostly due to absorption in the glass substrate, no sign change is observed. The measurements on all other samples showed similar results. Hence, the experiments have revealed no indication that the emitted THz radiation is due to a spin-current or similar symmetry breaking effects. From all these features we conclude that ultrafast laser-induced demagnetization can be reliably assigned to be the main source of the observed THz emission from the studied samples.

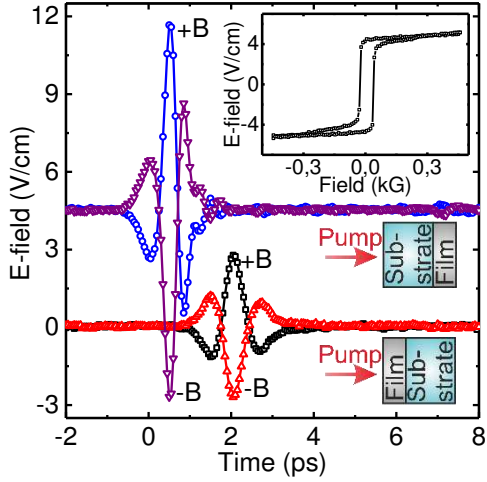

FIG. 2. (color online). Electric field of the emitted THz radiation from the $Nd_{0.2}(Fe_{0.87}Co_{0.13})_{0.8}$ sample as a function of time. The measurements have been performed in an external magnetic field of 1 kG and -1 kG (+ B and – B). The waveforms are measured for two orientations of the sample as indicated next to curves. The inset shows the hysteresis of the peak amplitude when the pump pulse excites the sample from the side of the substrate. The position of zero time for the upper curves is arbitrary chosen. The additional delay of the data shown in the lower curves is due to different propagation speeds of the THz and visible radiation inside the substrate.

To compare the THz emission and the MOKE measurements, we assumed that the dynamics of the MOKE is fully defined by the dynamics of the net magnetization. In the simplest approximation, the ultrafast demagnetization emission and magneto-optical probe data are compared with each other assuming coherent excitation of magnetic dipoles and observing them in far-field [20]. This gives $E \propto \frac{\partial^2 M}{\partial t^2}$, where $E$ the electric field of the emitted THz radiation in the time-domain, $M$ is the net magnetization in time-domain and $t$ is time. This expression, however, disregards any effect of optical components on the propagation of the radiation and significant size of the source, which deform the observed terahertz radiation significantly both in time and space. To derive a more precise solution, we solve the Maxwell equations with corresponding boundary conditions for an infinite film with homogenous magnetization dynamics. We use Faraday's and Ampère's laws to come to an expression for the spectrum of the electric field (in Gaussian units) generated at the surfaces of the ferromagnet:

$$\tilde{E}_y = \frac{4\pi\omega}{c} i\tilde{m}_x d \frac{1}{n+1} e^{-i\frac{\omega}{c}nz}, \qquad (1)$$

where $\tilde{E}_y$ is the y-component of the electric field of the THz emission in the frequency domain, $\omega$ is the angular frequency, $c$ the speed of light, $\tilde{m}_x$ the x-component of the magnetization in the frequency domain, $d$ the thickness of the magnetic layer and $n$ the refractive index of the substrate. For these calculations we used information about the net magnetization of the alloys in thermodynamic equilibrium. The derivation of Eq. 1 as well as details of the calculations of the anticipated spectra of the THz emission from the time-dependencies of the MOKE are presented in the supplementary information.

In Fig. 3 we compare our measured spectra (solid lines) with spectra calculated from the time-resolved MOKE data (dashed lines) for different samples. For the calculated spectra of Co we assumed that the net magnetization is equal to 1400 emu/cm$^3$ [24]. For $Gd_{0.3}(Fe_{0.87}Co_{0.13})_{0.7}$ we took the magnetization equal to 150 emu/cm$^3$ [25]. For $Nd_{0.2}(Fe_{0.87}Co_{0.13})_{0.8}$ we used 462 emu/cm$^3$ as measured with the help of a vibrating magnetometer and consistent with literature [26-25]. The spectra obtained for the Co sample show excellent mutual agreement. These observations show that THz emission and MOKE for a pure ferromagnetic material provide identical information of the ultrafast magnetization dynamics. Hence, regarding the totally different nature of the probes, such an agreement is a good indication of the viability of the two techniques for studying ultrafast magnetization dynamics.

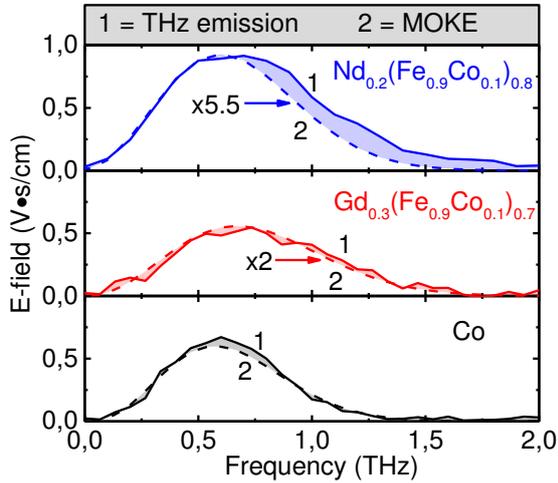

FIG. 3. (color online). Spectra of the THz emission. The solid lines (1) are spectra obtained with the help of the Fourier transform of the measured THz waveforms. The dashed lines (2) indicate the calculated spectra inferred from the time-resolved MOKE data. The calculated spectra of GdFeCo and NdFeCo are shown multiplied by 2 and 5.5 respectively. The shaded areas are meant to highlight the difference between the calculated and measured spectra.

For GdFeCo the spectrum calculated from the MOKE and the actually measured spectrum of the THz emission show a mismatch in amplitudes by a factor of 2. Simultaneously, we observe that the full-width at half maximum (FWHM) of the spectrum obtained using the MOKE agrees well with the FWHM of the actually measured THz spectrum (see Table 1). The expectation values of the spectra are also similar. Hence there is a satisfactory quantitative and a good qualitative agreement between the spectrum calculated from the MOKE and the actually measured spectrum of the THz emission during ultrafast laser induced demagnetization of GdFeCo.

Neither quantitative nor qualitative agreement between the spectrum calculated from the MOKE and the actually measured spectrum of the THz emission was obtained in the case of ultrafast laser-induced demagnetization of NdFeCo. In particular, Fig. 3 clearly indicates the lack of higher frequencies in the spectrum calculated from the MOKE response of NdFeCo, see also table 1. This implies that the dynamics revealed by the MOKE is slower than the magnetization dynamics revealed with the help of THz emission spectroscopy.

|  | Co | GdFeCo | NdFeCo |
|---|---|---|---|
| $FWHM_{MOKE}$ | 0.62 THz | 0.76 THz | 0.70 THz |
| $FWHM_{THz}$ | 0.58 THz | 0.73 THz | 0.81 THz |
| $|FWHM_{THz} - FWHM_{MOKE}| / FWHM_{MOKE}$ | 6.1 % | 4.4 % | 16 % |
| $mean_{MOKE}$ | 0.65 THz | 0.76 THz | 0.67 THz |
| $mean_{THz}$ | 0.64 THz | 0.79 THz | 0.77 THz |
| $|mean_{THz} - mean_{MOKE}| / mean_{MOKE}$ | 1.1 % | 4.2 % | 14 % |

Table 1. Full-width at half maximum (FWHM) and expectation value (i.e. center of mass) of the spectra presented in Fig. 3.

To determine the origin of the observed disagreements in the spectra, we note that the GdFeCo and NdFeCo samples have similar structures. Hence the layered structure of the studied samples cannot explain the qualitatively different results obtained for these two types of alloys.

One may argue that the observed discrepancies between the calculated and measured spectra for NdFeCo can be due to elemental specificity of the MOKE, which at the wavelength of 800 nm is mainly sensitive to the magnetization of the Fe-sublattice. Note that the THz emission is characterized by a different sensitivity, which follows the net magnetization dynamics. As the MOKE is mainly sensitive to the magnetization of the Fe-sublattice, the MOKE signals measured for $Gd_{0.18}(Fe_{0.87}Co_{0.13})_{0.82}$ and $Gd_{0.3}(Fe_{0.87}Co_{0.13})_{0.7}$ in an external magnetic field should have different signs. This sign change was indeed observed in the MOKE measurements, but not in the THz emission measurements (see supplementary information) revealing that at least the sign of the THz emission is defined by the net magnetization in these alloys.

However, if in NdFeCo the dynamics measured with the MOKE is associated with the dynamics of the FeCo sublattice, the mismatch in spectral distribution shows that the Nd-sublattice exhibits faster magnetization dynamics compared to the FeCo sublattice. This is in strong contrast to GdFeCo in which X-ray measurements revealed that the dynamics of the Gd sublattice in the GdFeCo alloy demagnetizes slower than the FeCo sublattice [16]. Analyzing the similarities and differences of GdFeCo and NdFeCo alloys, indicates that the orbital magnetization of Nd plays a significant role in the ultrafast demagnetization process. However, our experiments cannot resolve the exact roles of spin momentum, orbital momentum and spin-orbit interaction in the process of ultrafast laser induced demagnetization. Resolving this issue is an intriguing subject for future studies with the help of X-ray techniques.

To summarize we have combined pump-probe MOKE and THz spectroscopy techniques for simultaneous observation of the ultrafast magnetization dynamics and directly compare them with each other. For NdFeCo we observed clear differences in the MOKE and THz responses, showing that the MOKE measures slower dynamics than the magnetization dynamics that actually takes place.

We would like to thank T. Toonen and A. van Etteger for technical support. We would like to thank J. D. Costa for providing the Co sample, M. Huijben for the vibrating magnetometer measurements and J. Becker for fruitful discussions. This work was supported by the Foundation for Fundamental Research on Matter (FOM), the European Unions Seventh Framework Program (FP7/2007-2013) grant No. 280555 (Go-Fast), and European Research Council grant No. 257280 (Femtomagnetism).

**SUPPLEMENTARY INFORMATION:**

**Revealing the role of orbital magnetism in ultrafast laser-induced demagnetization in multisublattice metallic magnets by terahertz spectroscopy**

T. J. Huisman[1], R. V. Mikhaylovskiy[1], A. Tsukamoto[2], Th. Rasing[1] and A. V. Kimel[1]

[1]Radboud University Nijmegen, Institute for Molecules and Materials, 6525 AJ Nijmegen, The Netherlands
[2]College of Science and Technology, Nihon University, 7-24-1 Funabashi, Chiba, Japan

## I. THz emission derived from the Maxwell equations

Here we derive mathematical expressions describing how a fast laser induced demagnetization gives rise to the emitted THz radiation.

We assume a thin metal film is in the *xy* plane on top of a substrate as indicated in Fig. S1. The film is assumed to be magnetized along the x-axis and exhibits homogeneous magnetization dynamics. From Ampere's and Faraday's laws we can relate the magnetic field of the electromagnetic radiation to the magnetization of the medium as:

$$\frac{\partial}{\partial z}\left[\frac{1}{\varepsilon(z)}\frac{\partial \tilde{H}_x}{\partial z}\right] + \frac{\omega^2}{c^2}\tilde{H}_x = -\frac{4\pi\omega}{c^2}\tilde{M}_x, \quad \text{(S1)}$$

where $\tilde{H}_x$ is the x-component of the magnetic field in the frequency domain, $\varepsilon(z)$ is the dielectric permittivity, $\omega$ is the angular frequency, $c$ is the speed of light and $\tilde{M}_x = d\tilde{m}_x\delta(z)$ is the x-component of the magnetization in the frequency domain, $d$ is the thickness of the magnetic film and $\delta(z)$ is the Dirac function. We are looking for the plane wave solutions obeying to

$$\tilde{H}_x = \begin{cases} Ce^{i\frac{\omega}{c}z}, & z<0 \\ De^{-i\frac{\omega}{c}nz}, & z>0 \end{cases}. \quad \text{(S2)}$$

Here $n$ is the refractive index of the substrate. Continuity of $\tilde{H}_x$ implies $C=D$, while integration of Eq. S1 over the thickness of the film implies $\left[\frac{1}{\varepsilon(z)}\frac{\partial \tilde{H}_x}{\partial z}\right] = -\frac{4\pi\omega}{c^2}\tilde{m}_x d$, which results in $D = -\frac{4\pi\omega}{c}i\tilde{m}_x d\frac{n}{n+1}$. From Faraday's law it follows (in Gaussian units):

$$\tilde{E}_y = \frac{4\pi\omega}{c}i\tilde{m}_x d\frac{1}{n+1}e^{-i\frac{\omega}{c}nz}. \quad \text{(S3)}$$

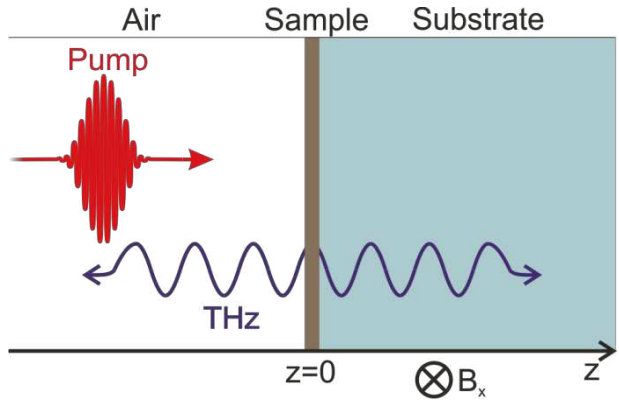

FIG. S1. (color online). Geometry of the modelled experiment. A laser pump pulse initiates magnetization dynamics in the sample which is accompanied by the emission of electromagnetic radiation.

To determine the generated electromagnetic radiation, the spectrum $\tilde{m}_x$ is required. The magnetization dynamics is inferred from time-resolved magneto-optical Kerr effect (MOKE) probe data. To prevent numerical errors the observed MOKE rotation is fitted with an empirical function, which is either:

$$\frac{\Delta M}{M} = \left(\frac{1}{2}\mathrm{erf}\left(A(t-B)+\frac{1}{2}\right)\right)\left(Ce^{-Dt} + Ee^{-Ft}\right), \qquad (S4)$$

or

$$\frac{\Delta M}{M} = \left(\frac{1}{2}\mathrm{erf}\left(A(t-B)+\frac{1}{2}\right)\right)\left(C\ln(Dt)e^{-Et}\right), \qquad (S5)$$

with $A$, $B$, $C$, $D$, $E$ and $F$ as fitting parameters. The choice of the fitting function depends on the shape of MOKE rotation in time. Figure 1 (b) in the main text of the paper shows the magneto-optical probe data fitted using Eqs. S4 and S5. Figure S2 shows the electromagnetic radiation emitted as a result of the magnetization dynamics inferred from the time-resolved MOKE data.

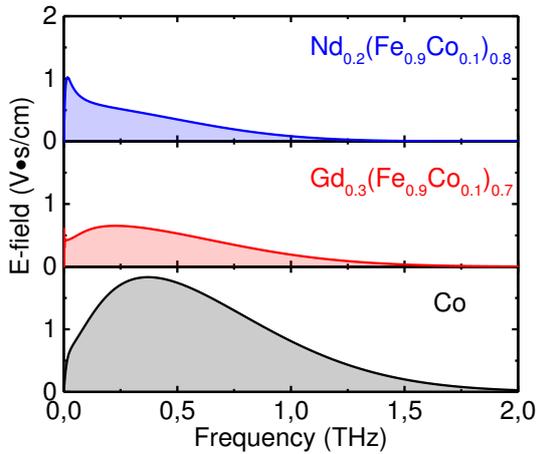

FIG. S2. (color online). The calculated spectra of the emitted electromagnetic radiation inferred from the time-resolved magneto-optical data shown in Fig. 1 (b) of the main text of the paper.

## II. Influence of the spectrometer on the generated THz emission

The observed spectrum of the emitted electromagnetic radiation can be related to the spectrum of the generated radiation by a linear relation:

$$E_{observed} = K(\omega) E_{source}, \qquad (S6)$$

where $K(\omega)$ is the transfer function which accounts for the substrate transmission, propagation effects and response of the ZnTe crystal. The transmission through a substrate much thicker than the wavelength of the electromagnetic wave is given by the Fresnell transmission equation:

$$t = \frac{2 n_i}{n_i + n_t} \qquad (S7)$$

where $n_i$ is the refractive index of the medium from which the radiation originates and $n_t$ is the refractive index of the medium to which the radiation is transmitted. The refractive index of the glass substrate is obtained experimentally with the help of THz transmission spectroscopy as described in [1]. Our results are comparable with the observations in [2]; the absorption of glass can be approximated by an increasing quadratic function with increasing THz frequencies, which effectively suppresses high frequencies of the emitted radiation.

To come from a near-field solution to the THz radiation at the detecting crystal, we apply Gaussian propagation similar to the one described in [3]. In order to apply Gaussian propagation, one needs to provide an initial diameter for the Gaussian beam in the model. Our initial diameter used in the calculations is taken equal to the diameter of the pump beam at the sample. In our setup, the effects of propagation remove low frequencies in the spectrum. Frequencies above 2 THz are all enhanced by a factor equal to the ratio of the focal lengths of the two parabolic mirrors used to collect and refocus the THz emission.

For the ZnTe response we applied the methods mentioned in [4], which shows that higher frequencies are suppressed. To compare quantitatively the calculated spectra with the experimentally obtained ones, we used Eq. 9 from [5] which shows how the observed ellipticity in the ZnTe crystal is related to the electric field amplitude of the THz radiation.

Taking into account both the propagation effects and the ZnTe response defines the spectrometer response, which can be visualized as a bandpass filter centered around 1.7 THz (see Fig. S3). The ZnTe crystal attenuates frequencies above 1.7 THz, while effects of propagation attenuate frequencies below 1.7 THz.

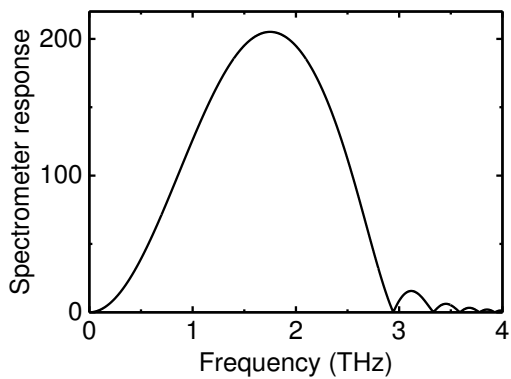

FIG. S3. The spectrometer response as function of frequency.

## III. Additional data for GdFeCo

The two used GdFeCo samples, $Gd_{0.18}(Fe_{0.87}Co_{0.13})_{0.82}$ and $Gd_{0.3}(Fe_{0.87}Co_{0.13})_{0.7}$, have different ratios between the magnetizations of the sublattices. For $Gd_{0.18}(Fe_{0.87}Co_{0.13})_{0.82}$ the magnetization of the FeCo sublattice dominates the net magnetization, while for $Gd_{0.3}(Fe_{0.87}Co_{0.13})_{0.7}$ the magnetization of the Gd sublattice dominates the net magnetization. Figure S4 (a) shows the dynamics of the MOKE measured for these two samples. It is seen that the dynamics have opposite signs, which is the result of different sublattices dominating the magnetization. Figure S4 (b) shows that the signs of the THz emission are the same for the two GdFeCo samples. The difference in amplitudes can be related to the difference in the net magnetizations.

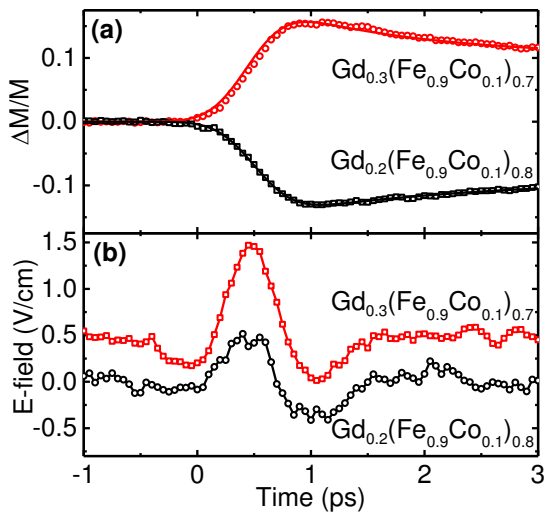

FIG. S4. (color online). (a) MOKE measured demagnetization for $Gd_{0.18}(Fe_{0.87}Co_{0.13})_{0.82}$ and $Gd_{0.3}(Fe_{0.87}Co_{0.13})_{0.7}$. The solid lines are fitting functions as expressed in Eq. S4. (b) Measured demagnetization emission of $Gd_{0.18}(Fe_{0.87}Co_{0.13})_{0.82}$ and $Gd_{0.3}(Fe_{0.87}Co_{0.13})_{0.7}$. The traces are vertically shifted with respect to each other. The position of zero-time is chosen arbitrary so that the position of the peak of the pump pulse is close to zero.